\documentclass[preprint,pre]{revtex4}
\usepackage{amssymb}
\usepackage{amsmath}
\usepackage{epsfig}

\def\expect#1{\langle#1\rangle}

\def\id3{ {\mathbf{1}_{3\times 3}}}

\def\e {\text{e}}
\def\D {\text{d}}
\begin{document}

\title{Hyperscaling relations in the bosonic pair contact process with diffusion
\\
}
\author{Matthias Paessens}
\email{m.paessens@fz-juelich.de}
\affiliation{Institut f\"ur Festk\"orperforschung, Forschungszentrum J\"ulich - 
52425 J\"ulich, Germany}
\date{\today}

\begin{abstract}
{
A hyperscaling relation for the critical exponents of absorbing phase transitions is
tested in the bosonic pair contact process with diffusion.
To this end spreading is considered, i.e. the time evolution out of an initial seed.
It is shown that like in the case of spatial homogeneous initial conditions the autocorrelation
function exhibits a phase transition at the critical point of the first moment. 
Some of the critical exponents can be determined exactly which is an unusual
property of an absorbing phase transition and provides a possibility to test
the hyperscaling relation. In the case of the bosonic pair contact process with diffusion
three sets of exponents can be considered referring to the number of particles,
number of pairs and number of active sites. 
It is argued that in special cases it is generally impossible to produce adequate 
data numerically.
}

\end{abstract}

\maketitle
\section{Introduction}

A possibility to investigate non--equilibrium critical phenomena
is given by absorbing phase transitions. These are transitions
from an active fluctuating
phase with a finite particle density to an absorbing state where any dynamics
is suppressed.
A central question in the theory of critical phenomena is the determination of
the universality class of a given system. 
In the field of absorbing phase transitions rather robust universality classes
have been found, e.g. the class
of directed percolation (DP) and the parity conserving universality class (PC).
For a review see Ref.~\cite{Hinr00}.

In order to determine the respective universality class the calculation of 
critical exponents plays a central role.
One possibility to determine critical exponents is to consider the scaling
of stationary quantities with respect to the distance from the critical point
$\Delta=p-p_c$ where $p$ is the control parameter and $p_c$ the critical point.
One defines the exponents $\beta, \beta^\prime, \nu_\bot, \nu_\|$ by
\begin{equation}
\begin{split}
\rho^\text{stat} &\propto \Delta^\beta, \\
P_\infty&\propto \Delta^{\beta^\prime},\\
\chi_\bot &\propto \vert\Delta\vert^{\nu_\bot},\\
\chi_\| &\propto \vert\Delta\vert^{\nu_\|},
\end{split}
\end{equation}
where $\rho^\text{stat}$ is the stationary particle density, 
$P_\infty$ is the ultimate survival probability, i.e. the probability that a randomly
chosen site
belongs to an infinite cluster, and 
$\chi_\bot, \chi_\|$ are the spatial and temporal correlation lengths.

Another possibility is given by dynamical scaling where the time dependence
of the quantities when started from an initial seed
 is used to define the exponents. At the critical
point, $\Delta=0$, one defines:
\begin{equation}
\begin{split}
P(t)&\propto t^{-\delta}, \\
\rho(t)&\propto t^{-\alpha},\\
\expect{N(t)} &\propto t^{\theta},
\end{split}
\end{equation} 
where $P(t)$ is the probability that a system survives at time $t$, i.e. that
there are still active particles left, $\rho(t)$ is the particle density
{\em inside} an active cluster and $\expect{N(t)}$ is the particle number
averaged over {\em all}, i.e., active and inactive, systems. This set of exponents is not independent 
of the previous set, one can deduce generally for second order transitions \cite{Hinr00}
\begin{equation}
\delta=\beta^\prime/\nu_\|, \qquad \alpha=\beta/\nu_\|.
\end{equation}
Furthermore the following argument gives another relation between the exponents:
The particle density inside an active cluster is given by the particle
number in a specific {\em active} system, $N(t)$ divided by the spreading region $R(t)$
\begin{equation}
\rho(t)= \expect{ \frac{N(t)}{R(t)} }_\text{active},
\end{equation}
here $\expect{\cdot}_\text{active}$ indicates that the average is taken only over active
systems. For large times one expects that
\begin{equation}
\expect{ \frac{N(t)}{R(t)} }_\text{active}\propto \frac{ \expect{N(t)}/P(t)}{\expect{R(t)}}
\propto t^{\theta+\delta-d/z}, 
\end{equation}
as the spreading region scales like $t^{d/z}$ where $d$ is the dimension and $z=\nu_\|/\nu_\bot$
is the dynamical exponent.
Thus we get the hyperscaling relation 
\begin{equation}
\theta-d/z=-(\alpha+\delta).
\label{pcpd:eq:hyper}
\end{equation}

This rather intuitive derivation should hold for first \cite{Dick95} and second 
order phase transitions.
A more detailed derivation of the hyperscaling relation for 
second order transitions can be found in Refs.~\cite{Mend94,Hinr00}. 
It has also been shown that a hyperscaling relation can be defined for the case
of coupled systems, see Ref.~\cite{Kwon04} and references therein. 

As there is hardly any model used for investigating absorbing phase transitions that can be solved
analytically, all
critical exponents have to be determined numerically. An exception to this is the bosonic pair
contact process with diffusion (bosonic PCPD) for which a field theoretic approach
due to Howard and T\"auber \cite{Howa97} is available. In this context the term 'bosonic' refers
to the property that there is no exclusion rule for the particles,
each lattice site may be occupied by any number of particles. This leads to a theoretical
description in terms of bosonic operators instead of fermionic operators in the case of exclusion
models.

A drawback of this
model is that it exhibits a first order transition and is thus not suitable 
for deciding the universality class of the PCPD with particle number restriction, 
which is still an open problem \cite{Carl01,Bark03,Odor03,Odor00,Kock03}.  
For a comprehensive review of the 
current state of the art we refer to Ref.~\cite{Henk04b}.

However, an exceptional property of the bosonic PCPD is the fact that some of the
critical exponents are known exactly \cite{Paes04}. This information provides a possibility 
to test the commonly considered hyperscaling relation Eq.~(\ref{pcpd:eq:hyper}) 
for the critical exponents.
As still some quantities are not accessible analytically Monte Carlo
simulations are used to get the complete set of exponents. It turns out
that in some cases simulations of the bosonic PCPD are misleading in general.

\section{Model and formalism}
\label{pcpd:sec:model}
We consider the following process as introduced in Ref.~\cite{Howa97}: 
On a infinite $d$--dimensional cubic lattice particles ('$A$') are diffusing with rate $D$,
 in each spatial direction. Additionally
they branch and annihilate:
$k\ge 1$ particles $A$ are created with rate $\mu$ out of any set of $2$ particles, 
and $l\in\{1,2\}$ particles are annihilated with rate $\lambda$ out of any set of $2$ particles:
\begin{eqnarray}
2A &\overset{\mu}{\rightarrow}&(m+k)A \nonumber \\
2A &\overset{\lambda}{\rightarrow}&(p-l)A \nonumber \\
A\cdot & \overset{D}{\leftrightarrow}&\cdot A.
\end{eqnarray}
The number of particles on each lattice site is not restricted -- the creation and 
annihilation processes take place on one lattice site. Thus the bosonic representation
of the process is used. 
One special case is the PCPD, where $l=2$ and $k=1$.

For $\lambda l > \mu k$ the particles die out  according
to a power law, while for $\lambda l < \mu k$ the particle density 
diverges. 
In analogy to the exclusion model we call the
rate which divides the two different behaviors the ``critical'' rate, 
\begin{equation}
\lambda_c=\mu k /l
\end{equation}
for given $\mu$ \cite{Paes04}. 
For this rate the total particle number
is constant for all times. 

For the details of the formalism we refer to Refs.~\cite{Howa97,Paes04}. The time evolution
of the particle annihilation operators $a({\bf x})$ is derived for the case
$\lambda=\mu k / l$:
\begin{equation}\begin{split}
\frac{\partial}{\partial t} \expect{a({\bf x})} =& D \sum_{k=1}^{d} 
      \left\{     \expect{a({\bf x}-{\bf k})} + \expect{a({\bf x}+{\bf k})}
                -2\expect{a({\bf x})} 
      \right\}  \\
\frac{\partial}{\partial t} \expect{a({\bf x})a({\bf y})} 
                                          \underset{{\bf x}\ne {\bf y}}{=} &
  D \sum_{k=1}^{d} \left\{\right. \expect{ a({\bf x})a({\bf y}-{\bf k}) } + 
                           \expect{ a({\bf x})a({\bf y}+{\bf k}) } +\\
                 &         \expect{ a({\bf x}-{\bf k})a({\bf y}) } +
                           \expect{ a({\bf x}+{\bf k})a({\bf y}) } -
                         4 \expect{ a({\bf x})a({\bf y})} \left.\right\} \\
\frac{\partial}{\partial t} \expect{\left(a({\bf x})\right)^2} =& 
     2D \sum_{k=1}^{d} \left\{ \expect{ a({\bf x})a({\bf x}-{\bf k}) } +
                               \expect{ a({\bf x})a({\bf x}+{\bf k}) } -
                              2\expect{ a({\bf x})^2} \right\} \\
& +\mu k (k+l) \expect{a({\bf x})^m} 
\label{eq.5}
\end{split}\end{equation}
where ${\bf k}\equiv {\bf k}(k)=(\ldots,0,1,0,\ldots)^T$ is the $k$-th unit space vector.
The particle density is given by $\rho({\bf x},t)=\expect{n({\bf x})} =\expect{ a({\bf x}) }$ and the 
correlation function by $\expect{n({\bf x})n({\bf y})} = \expect{ a({\bf x})a({\bf y}) }
+ \delta_{\bf x,y} \expect{a({\bf x})}$.

\section{Activity spreading}
\label{pcpd:sec:spreading}
In the theory of absorbing phase transitions beneath spatial homogeneous initial
conditions often
the following scenario is used: Initially the lattice is empty except for
the origin where just as many particle are located as needed for the dynamics
to start. It is then investigated how this activity spreads into the system.
In this section the question is addressed what can be learned from this
initial condition in the bosonic PCPD. 

After presenting analytically solvable cases we test the hyperscaling
relation of the critical exponents. To this end additional information 
is needed which can be obtained from Monte Carlo simulations. 

\subsection{Analytical calculation}
\label{pcpd:subsec:analytical}
As initial condition of the system we choose two particles at the origin
\begin{equation}
\rho({\bf x},t=0) = 2 \delta_{\bf x,0}.
\label{pcpd_II:initial}
\end{equation}

For the average particle number $\expect{\sum_{\bf x} a({\bf x})}$ we recover the same
result as for the density in the case of spatially homogeneous initial conditions:
for $\lambda l < \mu k$ the particle number diverges (``active phase'')
 while all particles die out for $\lambda l > \mu k$ (``inactive phase''). In the active phase
a spreading cone forms, i.e. a growing region with non--zero density.
If $\lambda=\mu k /l$ is chosen
the time evolution of the density is simply a lattice diffusion equation 
and the solution is given by
\begin{equation}
\rho({\bf x},t) = 2 \e^{-2dt} I_{x_1}(2t)\cdot\ldots\cdot I_{x_d}(2t).
\end{equation}
For large arguments ${\bf x}$ and $t$ this function asymptotically approaches a Gaussian
distribution, thus the dynamical exponent is $z=2$ and 
the average particle number is constant, $\theta=0$.

By defining
\begin{equation}
F_{\bf x}({\bf r},t) = \expect{a({\bf x})a({\bf x+r})}
\end{equation}
and rescaling time by 
\begin{equation}
t\to\frac{t}{2D}
\label{pcpd_II:res_time}
\end{equation}
the time evolution of the second moment can be rewritten as
\begin{equation}
\begin{split}
F_{\bf x}({\bf r},t)=&\frac{1}{2} \sum_{k=1}^{d}\left\{ F_{\bf x}({\bf r-k},t)+F_{\bf x}({\bf r+k},t)\right. \\
& \left.+ F_{\bf x-k}({\bf r+k},t)+F_{\bf x+k}({\bf r-k},t)-4 F_{\bf x}({\bf r-k},t)\right\} \\
& + \delta_{\bf r,0} F_{\bf x}({\bf 0},t),
\label{pcpd_II:F_DGL}
\end{split}
\end{equation}
where 
\begin{equation}
\alpha=\frac{\mu k (k+l)}{2D}.
\end{equation}
We rescaled time by the factor $1/(2D)$ instead of $1/D$ in order to keep the notation consistent
with the previous publication, Ref.~\cite{Paes04}. 

Equation (\ref{pcpd_II:F_DGL}) can be solved using a two--component Fourier--trans\-for\-mation
\begin{equation}
\begin{split}
f({\bf s,q},t)=&\sum_{\bf x}\sum_{\bf r} \e^{-i{\bf sx}}\e^{-i{\bf qr}} F_{\bf x}({\bf r},t), \\
F_{\bf x}({\bf r},t)=&\int\,\frac{\D^d{\bf s}\,\D^d{\bf q}}{(2\pi)^{2d}}\e^{i{\bf sx}}\e^{i{\bf qr}} f({\bf s,q},t).
\end{split}
\end{equation}

The differential equation for the Fourier--transform $f$ can be cast into the form
\begin{equation}
\frac{\partial}{\partial t} f({\bf s,q},t)=- \frac{1}{2} v({\bf s,q}) f({\bf s,q},t) 
+ \alpha \hat{F}({\bf s},t),
\end{equation}
with the dispersion relation $v({\bf s,q})=-\sum_{k=1}^{d} \left(\cos(q_k)+\cos(q_k-s_k)-2\right)$ 
and
\begin{equation}
\hat{F}({\bf s},t)=\sum_{\bf x} \e^{-i\bf sx} F_{\bf x}({\bf 0},t) = \int \frac{\D^d{\bf q}}{(2\pi)^d} f({\bf s,q},t).
\end{equation}
Integration yields
\begin{equation}
f({\bf s,q},t)= \e^{-v({\bf s,q})t} \left\{f({\bf s,q},0)+ \alpha \int_0^t \D\tau
\hat{F}({\bf s},\tau) \e^{v({\bf s,q})\tau}\right\}.
\end{equation}
The initial condition equation (\ref{pcpd_II:initial}) reads in Fourier space
$f({\bf s,q},0)=2$. Thus we get for the correlation function
\begin{equation}
F_{\bf x}({\bf r},t)= 2 A({\bf x,r},t) 
+ \alpha\sum_{x^\prime} \int_{0}^{t}\D\tau F_{x^\prime}({\bf 0},\tau)A({\bf x-x^\prime,r},t-\tau),   
\end{equation}
where
\begin{equation}
A({\bf x,r},t)=\int\,\frac{\D^d{\bf s}\,\D^d{\bf q}}{(2\pi)^{2d}} \exp\left(-v({\bf s,q})t+ i {\bf sx}+i{\bf qr}\right).
\end{equation}

An analytical solution of this integral equation could not be found
but for the sum of the autocorrelations,
\begin{equation}
\hat{F}({\bf 0},t)= \sum_{\bf x} F_{\bf x} ({\bf 0},t),
\end{equation}
the situation simplifies because of the following identity
\begin{equation}
b(t):=\sum_{\bf x} A({\bf x-x^\prime,0},t)=\sum_{\bf x} A({\bf x,0},t)=
\int \,\frac{\D^d{\bf s}}{(2\pi)^{d}} \e^{-v({\bf 0,q}) t} = \left(e^{-2t} I(2t)\right)^d
\end{equation}
where $I$ is a modified Bessel function.
We thus get
\begin{equation}
\hat{F}(0,t)=2 b(t) + \alpha \int_0^t\,\D\tau \hat{F}(0,\tau) b(t-\tau).
\end{equation}
This equation is exactly the one for the Lagrangian multiplier in the mean spherical model and
we can use the already known results here \cite{Godr00b}.
Using a Laplace transformation,
\begin{equation}
\tilde{F}(p)=\int_{0}^{\infty}\!\D t\, \e^{-pt} \hat{F}({\bf 0},t),
\end{equation}
we get
\begin{equation}
\tilde{F}(p)=\frac{2 \tilde{b}(p)}{1-\alpha \tilde{b}(p)}.
\label{pcpd_II:laplace_solution}
\end{equation}
An analysis of the behavior for small $p$ gives the late time behavior of $\hat{F}({\bf 0},t)$.
Depending on the dimension we find a phase transition with respect to $\alpha$. 
The critical point is given by the same $\alpha_c$ as found before for spatially homogeneous
initial conditions as calculated in Ref.~\cite{Paes04}. 
Above the critical point, $\alpha>\alpha_c$,
 the sum of the autocorrelations diverges as before. At the critical point $\alpha=\alpha_c$
the sum of the autocorrelations follows a power law 
$t^{-(2-d/2)}$ for $2<d<4$
and approaches a constant for $d>4$. Below the critical point $\alpha<\alpha_c$ the sum of 
the autocorrelations follows a power law $t^{-d/2}$. Not that for $d=1$ the critical
point is zero, $\alpha_c=0$ \cite{Paes04}.

The behavior below the critical point can be understood using the interpretation from spatial
homogeneous initial conditions 
that in this regime the diffusion is dominant. A diffusive system without reactions shows
the same late time behavior of the autocorrelator:
\begin{equation}
\begin{split}
\expect{\left(n({\bf x},t)\right)^2}&=\sum_{n} n^2 p_n({\bf x},t)\\
&=\sum_{n} n^2 \left(P({\bf x},t\vert{\bf 0},0)\right)^n \\
\Rightarrow \sum_{\bf x}&\expect{\left(n({\bf x},t)\right)^2}-\expect{n({\bf x},t)}= \\
&=\sum_{\bf x}\sum_{n=2} n^2 \left(\e^{-2dt} I_{x_1}(2t)\cdot\ldots\cdot I_{x_d}(2t)\right)^n\\
&=\sum_x 4\int\frac{\D^d{\bf q}\,\D^d{\bf q^\prime}}{(2\pi)^{2d}} \exp\left(i {(\bf q+q^\prime)x}-
\left(w({\bf q})+w({\bf q^\prime})\right)t\right)  +\ldots\\
&=4\int\frac{\D^d{\bf q}}{(2\pi)^{d}} \exp\left(-2w({\bf q})t\right)+\ldots \\
&\underset{t\to\infty}{=} 4 \left(8\pi t\right)^{-d/2},
\end{split}
\end{equation}
where $p_n({\bf x},t)$ is the probability to find $n$ particles at site ${\bf x}$ at time $t$
which for independent particles starting from the origin can be expressed as products of the
propagator $P({\bf x},t\vert{\bf 0},0)$. 

Above the critical point the reaction processes are dominant. As on average the particle
number in each system is $2$ in most of the systems the particles have to vanish in order that
in few systems a divergence of the second moment is possible. This has crucial
influence on the possibility to simulate the process as discussed in the
next section. 

\subsection{Hyperscaling relation}
\label{pcpd:subsec:hyper}
%%----------------------------------------------------------------------------%%
\begin{figure}[tn]
\vspace{3mm}
\centerline{\epsfxsize=3.5in\epsfbox
{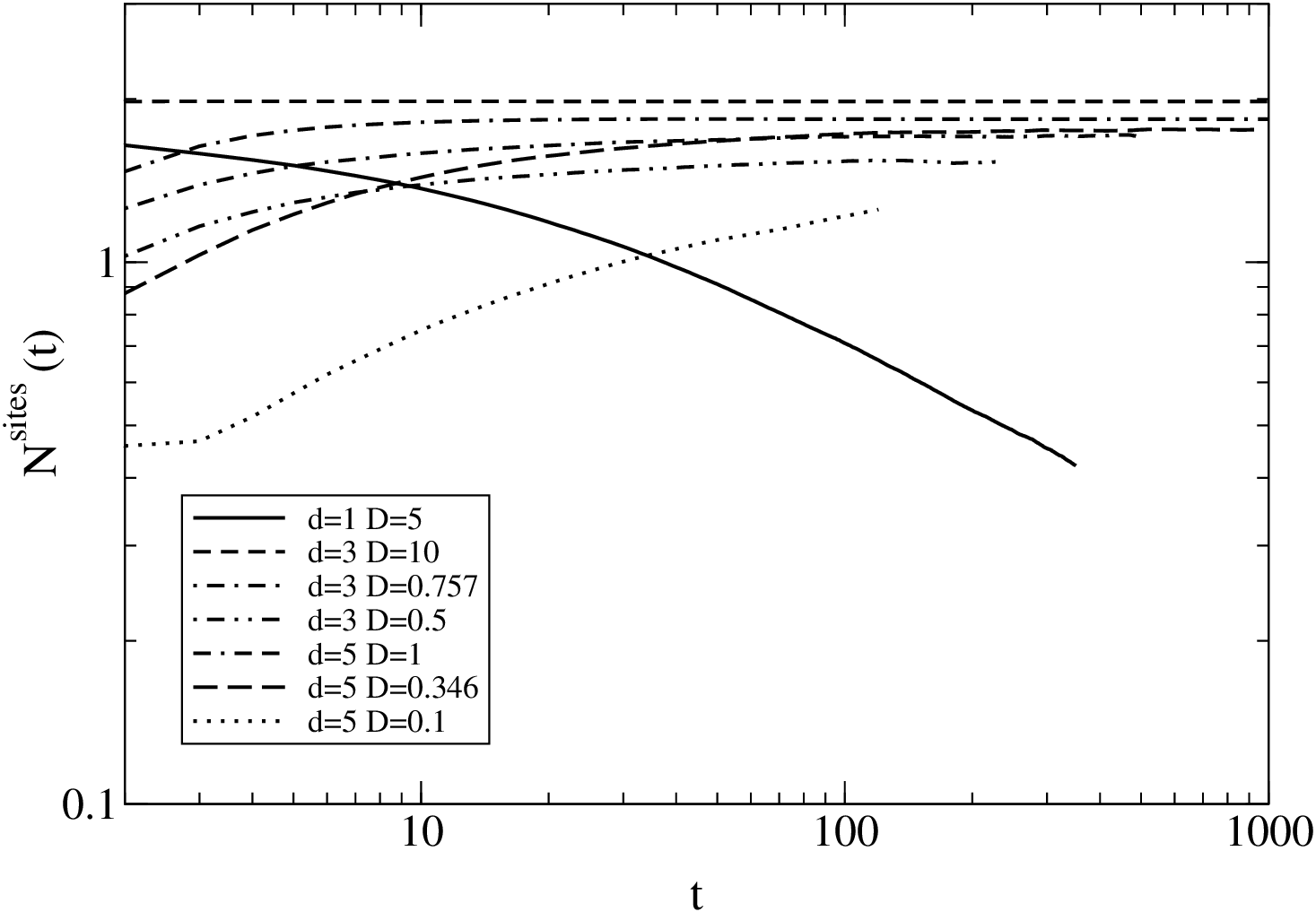}
}
\caption{Time evolution of the average number of active sites for different parameters, the reaction
rates are fixed to $\mu=2$ and $\lambda=1$. For $d=1$ always $\alpha>\alpha_c$, for $d=3$ and
$d=5$ the diffusion constant $D$ is chosen such that $\alpha<\alpha_c$, $\alpha=\alpha_c$ and
$\alpha>\alpha_c$ (with decreasing $D$). One observes that only in one dimension the number
of active sites does not approach a constant, a fit yields $\theta^\text{sites}_{d=1}\approx -0.450$.}
\label{pcpd:fig:N_sites}
\end{figure}
%%----------------------------------------------------------------------------%%%
%
\subsubsection{Analytical predictions}
%%----------------------------------------------------------------------------%%
\begin{figure}[tn]
 \vspace{3mm}
\centerline{\epsfxsize=3.5in\epsfbox
{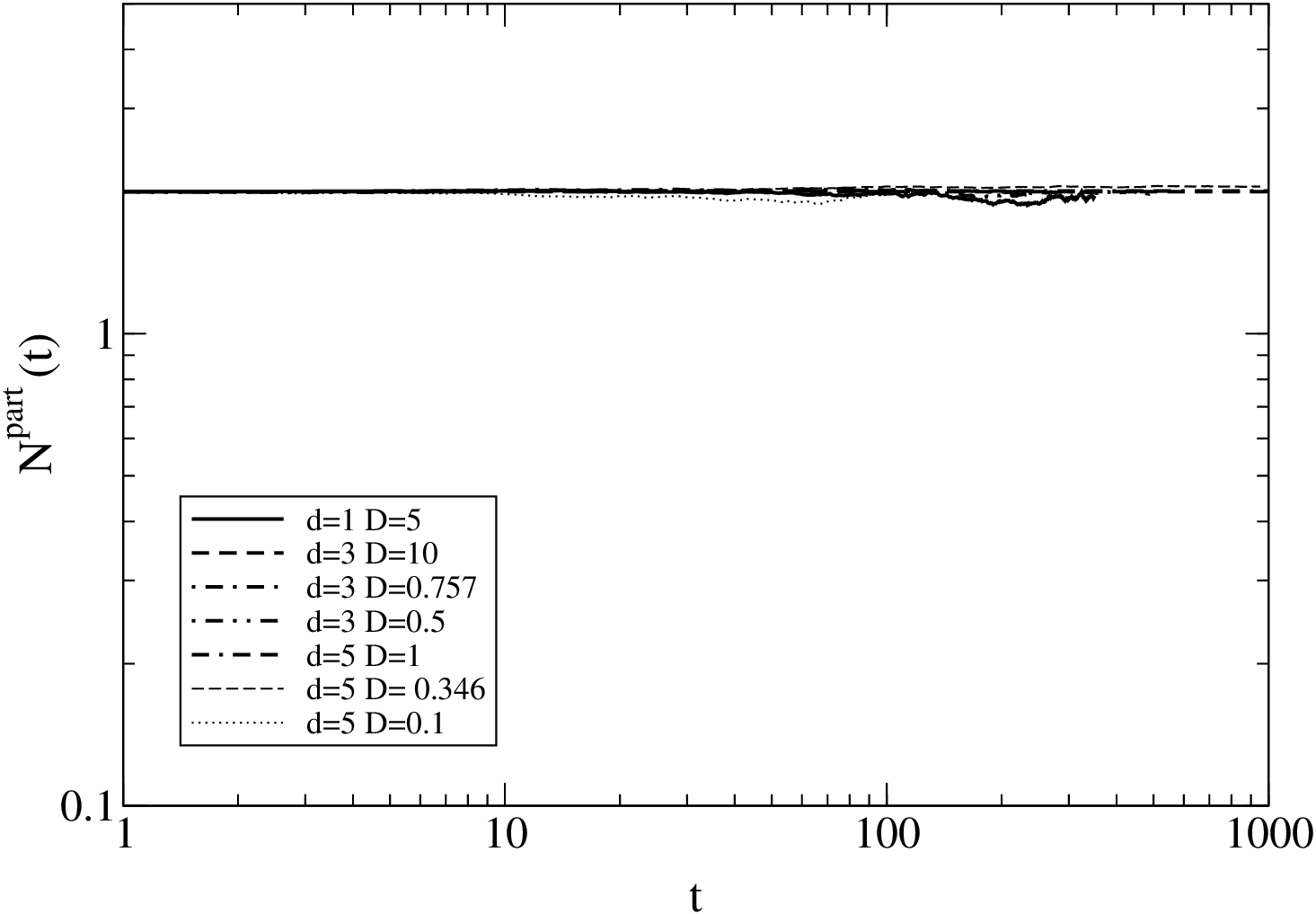}
}
\caption{Time evolution of the average number of particles, parameters as before. As expected from the
analytical calculations the number of particles is constant for any set of parameters.}
\label{pcpd:fig:N_part}
\end{figure}
%%----------------------------------------------------------------------------%%%
%

%%----------------------------------------------------------------------------%%
\begin{figure}[tn]
\vspace{3mm}
\centerline{\epsfxsize=3.5in\epsfbox
{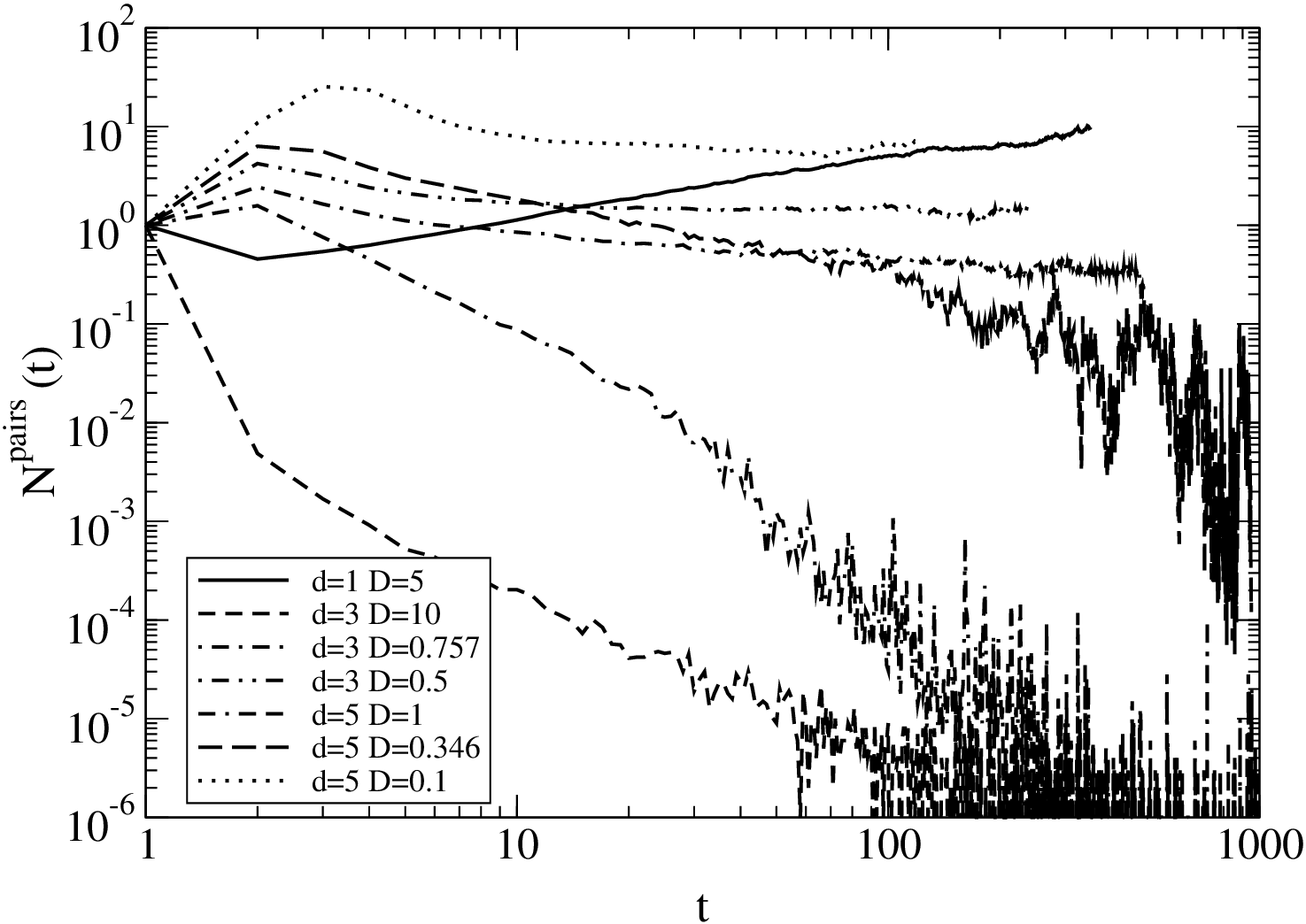}
}
\caption{Time evolution of the average number of pairs, parameters as before. In contrast to the
analytical result one can not observe the divergence for $\alpha>\alpha_c$ as explained in the text.
For $\alpha=\alpha_c$ and $\alpha<\alpha_c$ the fluctuations are still very large and the agreement
of the slopes with the analytically predicted values is not very high. Fitting yields
$\theta^\text{pairs}_{d=3}=-0.23$ to be compared to $-0.5$, $\theta^\text{pairs}_{d=5}=-0.76$ 
(analytically: $0$) for $\alpha=\alpha_c$ and $\theta^\text{pairs}_{d=3}=-1.7$ (analytically: $-1.5$),
$\theta^\text{pairs}_{d=5}\approx -2.5 (\pm 0.5)$ (analytically -2.5) for $\alpha<\alpha_c$. }
\label{pcpd:fig:N_pairs}
\end{figure}
%%----------------------------------------------------------------------------%%
% 

The hyperscaling relation Eq.~(\ref{pcpd:eq:hyper}) shall now be verified in the bosonic PCPD where we know
after all some of the exponents exactly. 
The arguments given for the hyperscaling relation should hold irrespectively
of the type of density/number which is measured. While in the description of the process
with exclusion interaction there are only two possibilities -- the number of particles
and pairs, i.e. two particles on neighboring sites -- in the bosonic description there 
are three possibilities. The number of active sites, these are sites with at least one particle,
have to be distinguished from the number of particles. So one may consider the number
of active sites, of particles or of particle pairs. The number of particles is constant at 
the critical point, thus we have $\theta^\text{part}=0$. The number of pairs is
given by  $\sum_{\bf x}\expect{(a^\dagger({\bf x}))^2 (a({\bf x}))^2}=\sum_{\bf x}\expect{a^2({\bf x})}$
whose behavior is calculated in subsection \ref{pcpd:subsec:analytical}:
\begin{equation}
\theta^\text{pairs} = \left\{\begin{array}{cl}
  -d/2     & \alpha<\alpha_c  \\
  -(2-d/2) & \alpha=\alpha_c \text{ and } 2<d<4 \\
  0        & \alpha=\alpha_c \text{ and } d>4. 
\end{array} \right.
\end{equation}
For $\alpha>\alpha_c$ the number of pairs diverges exponentially and thus no
exponent can be defined. 
%%----------------------------------------------------------------------------%%
\begin{figure}[tn]
\vspace{3mm}
\centerline{\epsfxsize=3.5in\epsfbox
{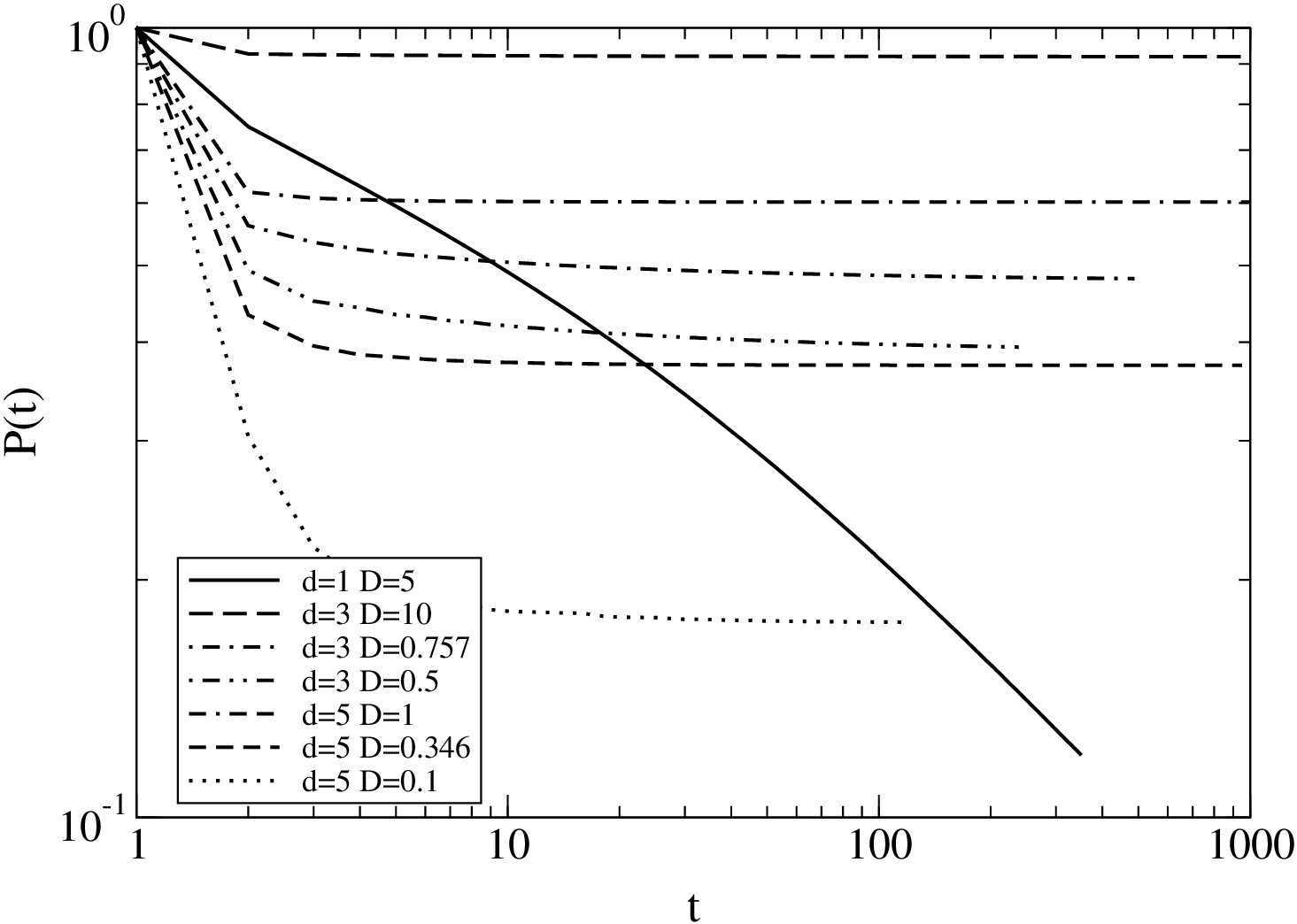}
}
\caption{Time evolution of the survival probability, parameters as before. As predicted, only
for $d=1$ the curve does not approach a constant, a fit yields in this case $\delta_{d=1}=0.471$.}
\label{pcpd:fig:delta}
\end{figure}
%%----------------------------------------------------------------------------%%
% 

In $d>2$ it is expected that $\delta=0$: After some initial time
there is a fraction of systems consisting only of two particles at different
sites. They are diffusing freely on the lattice and hence their distance
vector describes as well a random walk in $d$ dimensions. As a random walk
is transient for $d>2$ the probability that the two particles ever meet
again is zero and consequently those active systems survive for ever
and one concludes $\delta=0$.

\begin{table}[p]
\begin{tabular}{c||c||c||c||c|c}
    & & $\quad z \quad$   & $\quad\delta\quad$ & 
$\theta^\text{part}$ & $\theta^\text{pairs}$ \\
\hline
\hline
$d=1$ & $\alpha>\alpha_c$  & 2 & ? & 0 & -  \\
\hline
\hline
$2<d<4$  & $\alpha<\alpha_c$ & 2 & 0 & 0 & -d/2     \\
\hline 
$2<d<4$  & $\alpha=\alpha_c$ & 2 & 0 & 0 & -(2-d/2)  \\
\hline
$2<d<4$  & $\alpha>\alpha_c$ & 2 & 0 & 0 & -         \\
\hline
\hline
  $d>4$  & $\alpha<\alpha_c$ & 2 & 0 & 0 & -d/2      \\
\hline 
$d>4$    & $\alpha=\alpha_c$ & 2 & 0 & 0 & 0         \\
\hline
  $d>4$  & $\alpha>\alpha_c$ & 2 & 0 & 0 & -         
\end{tabular}
\caption{The exactly calculated exponents. Not defined exponents are represented by '-'
and the exponents to be determined in the simulations with '?'. The exponents 
$\theta^\text{sites}$, $\alpha^\text{sites}$, $\alpha^\text{part}$ and $\alpha^\text{pairs}$ 
have to be determined numerically.}
\label{pcpd:tab:ex_ex}
\end{table}
%%----------------------------------------------------------------------------%%%
%

Additionally we have shown exactly that $z=2$. The remaining exponents,
$\theta^\text{sites},\alpha^\text{sites},\alpha^\text{part},\alpha^\text{pairs}$
and $\delta_{d=1}$ have to be
determined numerically in a Monte--Carlo simulation. To sum up, the theoretical
considerations predict (see Eq.~(\ref{pcpd:eq:hyper}) and Tab.~\ref{pcpd:tab:ex_ex}):
\begin{equation}
\begin{split}
\alpha^\text{sites}&=\left\{\begin{array}{cl}
  -\theta^\text{sites}-\delta_{d=1}+\frac{1}{2}& d=1 \\
  -\theta^\text{sites}+\frac{d}{2}  & d>2 
\end{array} \right.\\
\alpha^\text{part} &=\left\{\begin{array}{cl}
  \qquad\frac{1}{2}-\delta_{d=1} \qquad& d=1 \\
  \frac{d}{2}  & d>2 
\end{array} \right.\\
\alpha^\text{pairs} &=\left\{\begin{array}{cl}
  \qquad\quad\,\,\, d \,\,\,\quad\qquad& \alpha<\alpha_c (d>2)\\
  2 & \alpha=\alpha_c \text{ and } 2<d<4 \\
  \frac{d}{2}  & \alpha=\alpha_c \text{ and } d>4 
\end{array} \right.
\end{split}
\end{equation}

%%----------------------------------------------------------------------------%%
\begin{figure}[tn]
\vspace{3mm}
\centerline{\epsfxsize=3.5in\epsfbox
{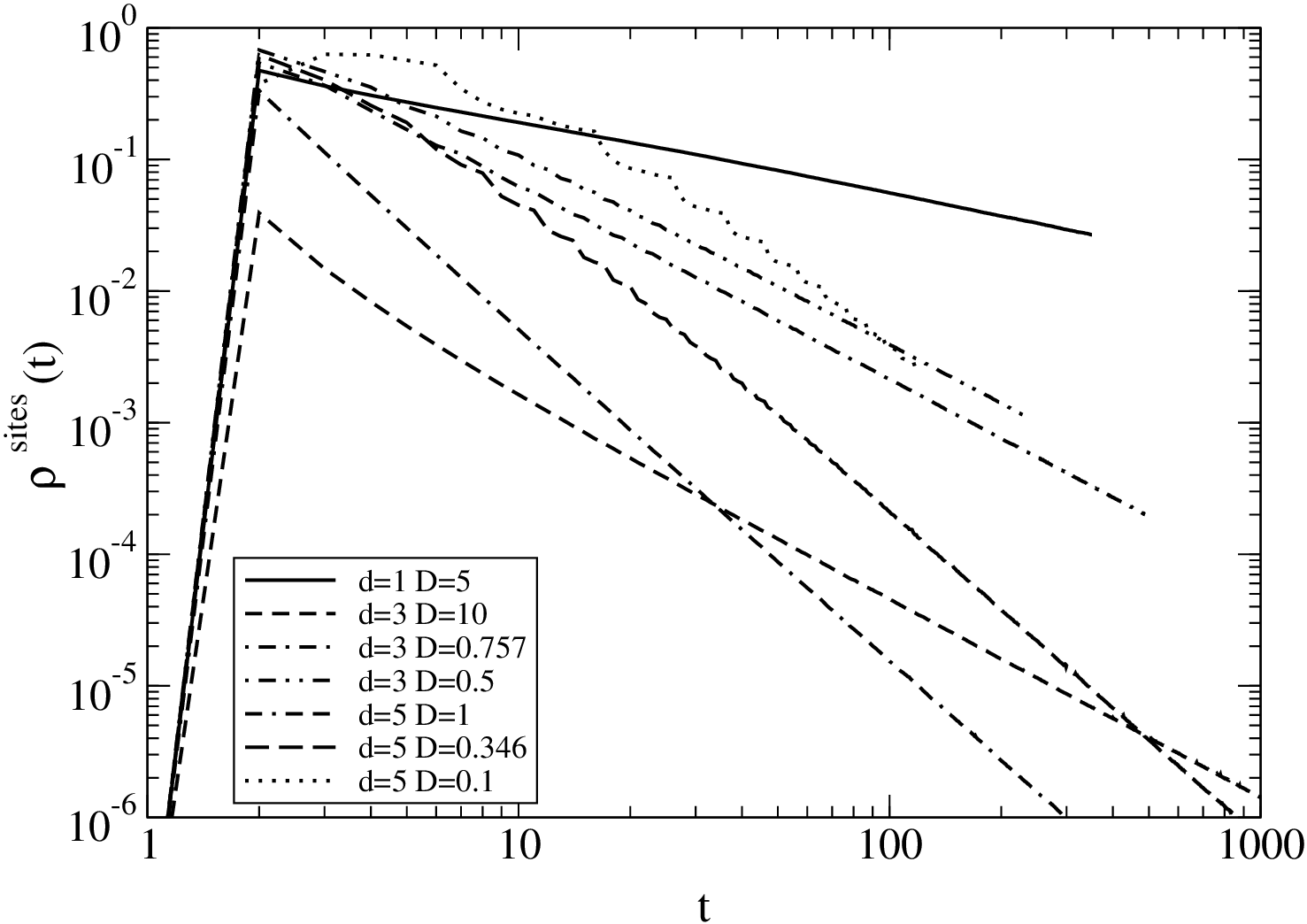}
}
\caption{Time evolution of the density of active sites, parameters as before. By fitting
these curves we observe that the exponents are given by $\alpha^\text{sites}=d/2$ in good
agreement, for exact values see table~\ref{pcpd:tab:exponents}.}
\label{pcpd:fig:rho_sites}
\end{figure}
%%----------------------------------------------------------------------------%%
% 
\subsubsection{Simulations}
For the Monte--Carlo (MC) simulation of the bosonic PCPD a list of active sites
$i$ is used. In contrast to the model with exclusion interaction the number of
particles $n_i$ on the sites has to be tracked. The number of pairs on each site
is then given by $n_i (n_i-1)/2$. For each system in one MC time
step $D \sum_i n_i$ diffusion processes, $\mu \sum_i n_i (n_i-1)/2$ creation
and $\lambda \sum_i n_i (n_i-1)/2$ annihilation processes take place on average.
One of these possibilities is chosen randomly according to its statistical weight
and the time is updated by 
$t\to t+\left[D \sum_i n_i+(\mu+\lambda) \sum_i n_i (n_i-1)/2\right]^{-1}$.
A difficulty is that the number of possible processes varies extremely from system
to system as the number of pairs fluctuates enormously. Consequently it is not
convenient to determine a target time and simulate each system up to this time
one after each other because for a badly estimated target time the program
might get stuck in only one of the systems with a large number of pairs.  
We rather determined target times in small steps up to which the systems were simulated
step  by step. Although the systems have to be kept simultaneously in memory
one gains the advantage that the results can be tracked during the simulations and one
does not have to estimate the maximal simulation time in advance.

%%----------------------------------------------------------------------------%%
\begin{figure}[tn]
\vspace{3mm}
\centerline{\epsfxsize=3.5in\epsfbox
{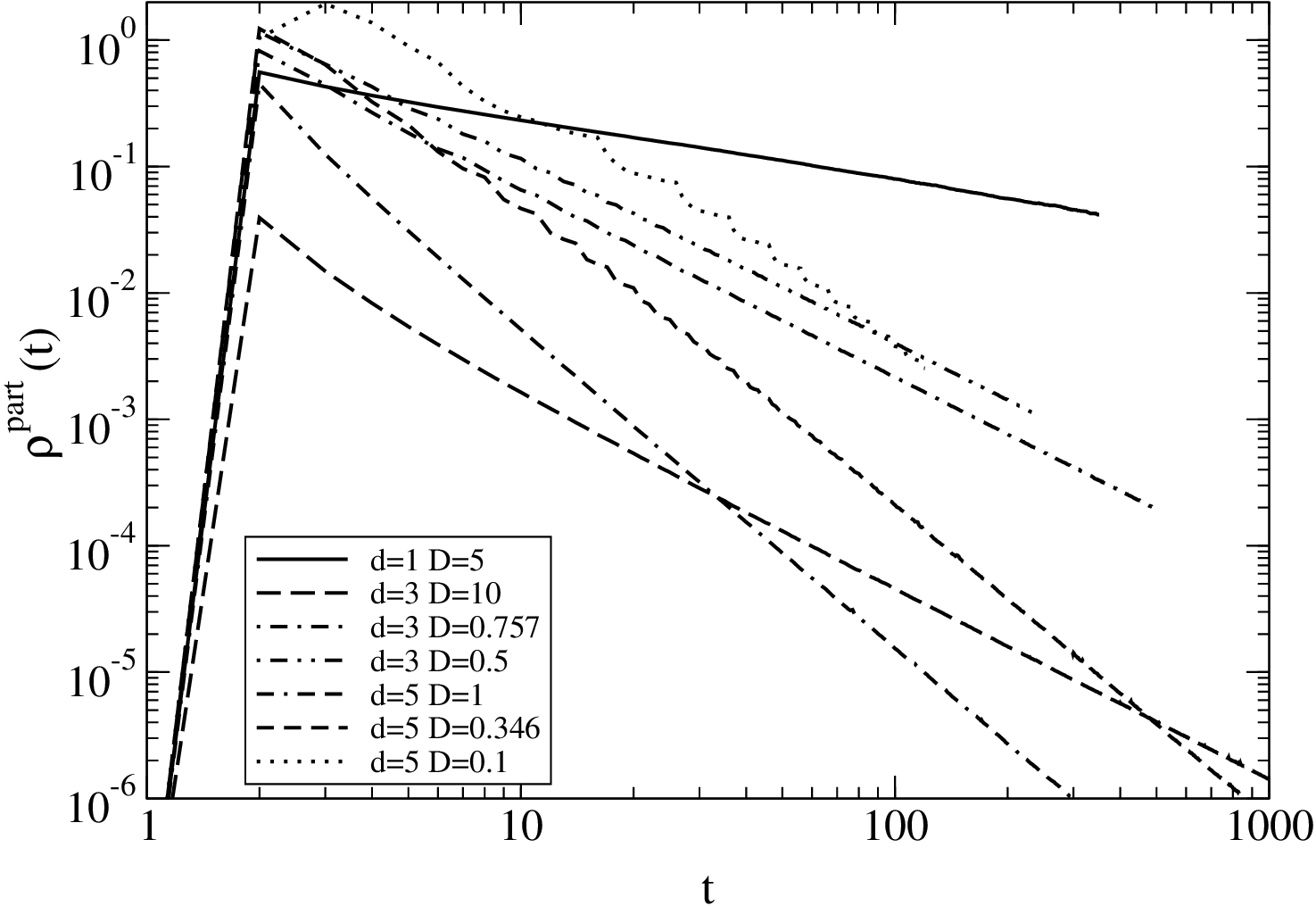}
}
\caption{Time evolution of the density of particles, parameters as before. By fitting
we again observe that the exponents are given by $\alpha^\text{part}=d/2$ in good 
agreement, for exact values see table~\ref{pcpd:tab:exponents}.}
\label{pcpd:fig:rho_part}
\end{figure}
%%----------------------------------------------------------------------------%%
% 
Still the simulation of the process takes much effort. Therefore compared to the simulations
of the model with exclusion interaction only small times could be simulated. Especially
in the case of diverging autocorrelations reliable results are computationally
demanding. Standard
simulation methods simply fail in this case as the number of needed systems in the ensemble is
far too large. This number of systems $M$ can be estimated as follows:
The average number of pairs diverges as
$\expect{N^\text{pairs}}\approx N^\text{pairs}_0 \exp(t/\tau)$ while the average number of particles 
$N_0$ is constant. To determine
a lower bound for $M$ one may assume that all the particles available in the simulation
$M N_0$ pile up at one site of a single system. 
Then the number of pairs is well approximated
by $(M N_0)^2/2$. This has to be equal to $M N^\text{pairs}_0 \exp(t/\tau)$ as this is the only
contribution to the ensemble average of the number of pairs. Thus we conclude 
that
\begin{equation} 
M>2 \frac{N^\text{pairs}_0}{N_0^2} \exp(t/\tau)
\end{equation}
systems are needed in order to allow for the divergence of autocorrelations.

In our simulations we have typically $\tau\approx 1$, for example for the
$3d$--case with $D=0.5,\mu=2,\lambda=1$ one gets $\tau=0.86$. Already for the simulation 
time $t=100$ one would need an ensemble consisting of roughly $10^{40}$ systems to observe
the divergence numerically. Consequently it is not expected that for $\alpha>\alpha_c$ 
(which is especially true in $d=1$) the simulation produces correct results.

%
%%----------------------------------------------------------------------------%%
\begin{figure}[p]
\vspace{3mm}
\centerline{\epsfxsize=3.5in\epsfbox
{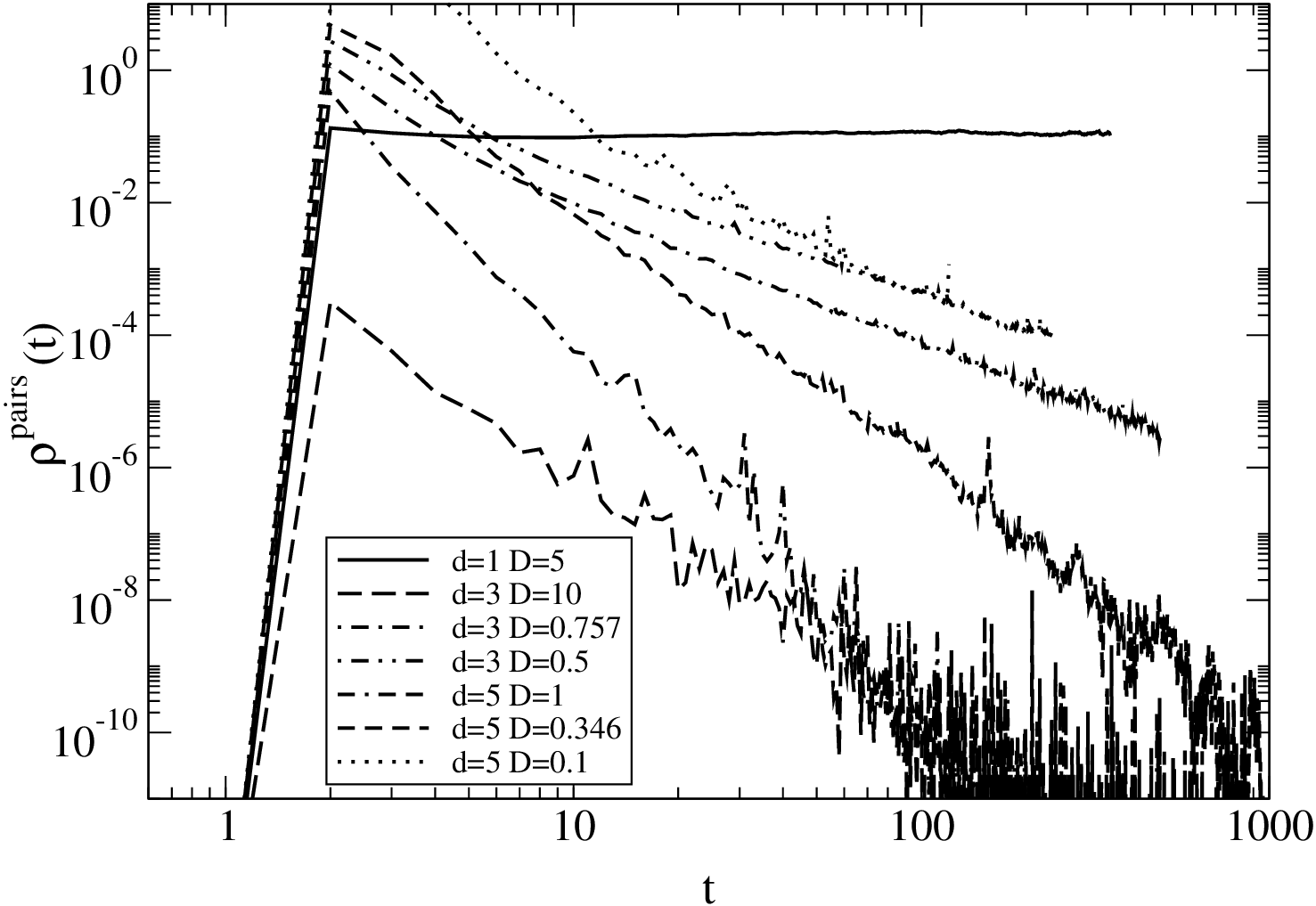}
}
\caption{Time evolution of the density of pairs, parameters as before.
Due to the fluctuations the accuracy of the exponent $\alpha^\text{pairs}$ is not very high. 
 A fit yields the values $\alpha^\text{pairs}_{d=3}=1.92$ 
to be compared to the theoretical value $2$, $\alpha^\text{pairs}_{d=5}=3.76$ to be compared
to $2.5$ for the case $\alpha=\alpha_c$ and $\alpha^\text{pairs}_{d=3}=3.58$ (analytically: $3$),
$\alpha^\text{pairs}_{d=5}\approx 6$ (analytically: $5$) for the case $\alpha<\alpha_c$. }
\label{pcpd:fig:rho_pairs}
\end{figure}
%%----------------------------------------------------------------------------%%
% 
%
%%----------------------------------------------------------------------------%%
\begin{table}[p]
\begin{tabular}{c|c||c|c||c|c||c|c||c|c}
$d$     & $D$   & $\delta$ & $\theta^\text{sites}$ &$\alpha^\text{sites}$ & $\alpha^\text{sites}_\text{calc}$ & $\alpha^\text{part}$ & $\alpha^\text{part}_\text{calc}$ & $\alpha^\text{pairs}$ & $\alpha^\text{pairs}_\text{calc}$  \\
\hline
\hline
1 & 5   & 0.471 & -0.450& 0.574 & 0.479 & 0.517 & 0.029 & -- & --\\
\hline 
3 & 0.5 & 0.009& 0.0108 &1.48 & 1.49 & 1.49& 1.5& -- & -- \\
\hline 
3 & 0.76 & 0.006 & 0.00429 & 1.49 & 1.49 & 1.49 & 1.5 & 1.92 & 2 \\
\hline
3 & 10  & 0.0002 & 2$\cdot$10$^{-5}$ & 1.50 & 1.5 & 1.50 & 1.5 & 3.58 & 3 \\
\hline
5 & 0.1  & 0.004 & 0.15 & 2.29 & 2.35 & 2.28 & 2.5 & -- & --\\
\hline 
5 & 0.35  & 5$\cdot$10$^{-5}$ & 0.0048 & 2.49 & 2.5 & 2.49 & 2.5 & 3.76 & 2.5\\
\hline
5 & 1  & $<$10$^{-10}$ & 3$\cdot$10$^{-5}$ & 2.51 & 2.5 & 2.51 & 2.5 & $\approx$ 6& 5   
\end{tabular}
\caption{The numerically determined values of the exponents compared to the values expected form
the hyperscaling relation (\ref{pcpd:eq:hyper}). For $\alpha>\alpha_c$ the exponent $\alpha^\text{pairs}$ is
not defined. It can be seen that for the number of sites and the number of particles the hyperscaling 
relation is satisfied in good agreement while for the number of pairs larger deviations appear.}
\label{pcpd:tab:exponents}
\end{table}
%%----------------------------------------------------------------------------%%%
%

Simulations with several parameters were performed: As dimensions $d=1$, 
$d=3$ and $d=5$ are chosen in order to simulate systems below the lower critical dimension,
above the lower but below the upper critical dimension and in the mean--field regime
of the phase transition of the second moment.
The reaction rates are fixed to $\mu=2$ and $\lambda=\lambda_c=1$ such that the average
number of particles remains constant in time. For $d=3$ and $d=5$ the diffusion constant $D$ is
chosen such that $\alpha<\alpha_c$ ($D=10$ for $d=3$ and $D=1$ for d=5), $\alpha=\alpha_c$
($D=0.757$ for $d=3$ and $D=0.346$ for $d=5$) and $\alpha>\alpha_c$ ($D=0.5$ for $d=3$ and
$D=0.1$ for $d=5$). In the ensemble $M=10^6$ systems are simulated in parallel.

Fig.~\ref{pcpd:fig:N_sites} shows the time evolution of the average number of active sites. For
$d=1$ the number of active sites decreases according to a power law with an exponent $\theta^\text{sites}_{d=1}=-0.45$.
For $d=3$ and $d=5$ for all parameters the curves approach a constant. As explained above this is expected
as a certain fraction of the systems will consist of at least two particles performing a random walk without
meeting again. 

In Fig.~\ref{pcpd:fig:N_part} the time evolution of the average number of particles is shown. In agreement
with the analytical calculations the particle number is constant in time and thus $\theta^\text{part}=0$
for all parameters. 

The time evolution of the average number of pairs is shown in Fig.~\ref{pcpd:fig:N_pairs}. Among the
three quantities considered this is clearly the most fluctuating one. The simulation indeed fails in
reproducing the exponential divergence of the number of pairs for $\alpha>\alpha_c$ as discussed
above. While for
$d=1$ the number of pairs is increasing according to a power law, for the higher dimensions
it approaches a constant. Interpreting the results for $\alpha>\alpha_c$ has thus to be done carefully.

The survival probability $P(t)$ is shown in Fig.~\ref{pcpd:fig:delta}. It is verified for $d=3$ and
$d=5$ that $P(t)$ approaches a constant while for $d=1$ it decays according to a power law. The fitted
values for $\delta$ can be found in table~\ref{pcpd:tab:exponents}. For $d=3$ and $d=5$ the values
are very close to zero and $\delta_{d=1}\approx 1/2$. 

The directly measured densities are shown in Fig.~\ref{pcpd:fig:rho_sites} (active sites),
Fig.~\ref{pcpd:fig:rho_part} (particles) and Fig.~\ref{pcpd:fig:rho_pairs} (pairs). 
For the densities of active sites and particles the fitted values for $\alpha$ are in
most cases in good agreement with the values obtained by the hyperscaling relation, only for
$d=1$ and $d=5, \alpha>\alpha_c$ larger deviations appear. As most of the exponents
take simple values, we conjecture in these cases that $\alpha^\text{sites}_{d=1}=1/2$ and $\alpha^\text{sites}_{d=5}=
5/2$ and that the measured deviations result from the numerical problems. 
An obvious disagreement between measured values and the hyperscaling
relation is found for $\alpha^\text{part}_{d=1}$, where the directly measured value is approximately
$1/2$ while the hyperscaling relation predicts it to be approximately zero. The question arises whether
the hyperscaling relation is violated or which of the values is wrong. As the hyperscaling relation
turns out to hold in the other cases we believe it to hold in this case as well
-- the inacuracy in the
MC results stems from the fact for $d=1$ the numerical problem is always present because
$\alpha_c=0$.
We conjecture that $\delta=1/2$ and $\alpha^\text{part}_{d=1}=0$. This can be imagined as follows: In the surviving systems the active regions spreads diffusively and inside the
active region the reaction kinetics generates a constant density.
Consequently the particle number increases in these
systems proportional to $t^{0.5}$, but as more and more systems die out according to $t^{-0.5}$
the particle number averaged over all systems remains constant. In $d=3$ and $d=5$ the situation
is different, the particle density inside the active region decays and the number of active
systems remains constant.

The values for $\alpha^\text{pairs}$ for $\alpha\le\alpha_c$ could not be determined with high accuracy
due to the high fluctuations. It would be surprising if the hyperscaling relation did not hold for
this quantity but the accuracy of our data allows neither for proving nor for disproving the
relation in this case.

\section{Conclusions}
\label{pcpd:sec:conclusion}
The evolution of the system out of an initial seed (activity spreading) is investigated in
the bosonic pair contact process with diffusion.
It is shown that the second moment exhibits a phase transition with the same
critical point as for spatially homogeneous initial conditions. Above the critical point the
sum of autocorrelations diverges and below the critical point it decreases according to a 
power law.
This power law behavior can be related to purely diffusive dynamics. This shows that 
below the critical point it can be neglected that the particles react because most
of the time the particles are diffusing freely. The time during which
two particles occupy the same lattice site is too short to react because  
below the critical point diffusion dominates above reactions. 

We tested a hyperscaling relation for the dynamical critical exponents. To this 
end exponents have to be determined which are not accessible analytically and are
thus calculated numerically in a Monte--Carlo simulation. It is shown that for
the case of a diverging second moment it is impossible to produce accurate
data as the necessary size of the ensemble diverges exponentially in the desired 
simulation time. At or below the critical point of the divergence of the second moment
the hyperscaling relation can be verified. It turns out that good accuracy
can be achieved for the number and density of active sites and the number and
density of particles while measuring these quantities for the number of pairs
is difficult due to large fluctuations. 

\section*{Acknowledgements}
G.M. Sch\"utz is gratefully acknowledged for suggesting this problem and many useful
discussions.

\end{document}